\documentstyle[twoside,fleqn,espcrc2]{article}
\input epsf
\def \beq{\begin{equation}}

\def \eeq{\end{equation}}

\def \G{{\rm GeV}}
\def \Hc{{\rm H.c.}}

\def \ite{{\it et al.}}

\def \k{K^0}
\def \bk{\bar K^0}

\def \M{{\rm MeV}}

\def \pipe{\pi^+ \pi^-}
\def \poop{\pi^0 \pi^0}

\def \s{\sqrt{2}}

\newcommand{\AmS}{{\protect\the\textfont2
  A\kern-.1667em\lower.5ex\hbox{M}\kern-.125emS}}

\title{CKM Matrix and Standard-Model CP Violation}

\author{Jonathan L. Rosner\address{Enrico Fermi Institute
        University of Chicago \\ 
        5640 S. Ellis Avenue, Chicago IL 60637}}
       
\begin{document}

\begin{abstract}
The currently favored model of CP violation is based on phases in the
Cabibbo-Kobayashi-Maskawa (CKM) matrix describing the weak charge-changing
couplings of quarks.  The present status of parameters of this matrix is
described.  Tests of the theory, with particular emphasis on the study of B
meson decays, are then noted.  Some remarks are made regarding the possible
origin of the baryon asymmetry of the universe; the corresponding coupling
pattern of the leptons could shed light on the question.  Some possibilities
for non-standard physics are discussed. 
\end{abstract}
\maketitle

\section{INTRODUCTION}

For more than thirty years, the neutral kaon system has been the only direct
place in which CP violation has been seen \cite{CCFT}.  The currently favored
theory of this phenomenon involves complex phases in the
Cabibbo-Kobayashi-Maskawa (CKM) matrix \cite{Cab,KM} describing the
charge-changing weak transitions of quarks.  It is very likely that a number of
experiments will be able to test this theory in the next few years.  The
present talk is meant as an overview of this activity, with particular emphasis
on $B$ decays.  A number of related subjects are covered in more detail by
other speakers at this Workshop.

We begin in Section 2 with a survey of the patterns of quark and lepton
masses and couplings.  The latter are described by parameters of the CKM
matrix whose current status we review in Section 3.  We then turn in Section 4
to some tests of this picture based on $B$ meson decays.  The baryon asymmetry
of the Universe, described briefly in Section 5, provides indirect evidence for
CP violation, though the CKM pattern alone is probably insufficent for
understanding it.  Some possibilities for non-standard physics are discussed in
Section 6, while Section 7 concludes. 

\section{QUARK AND LEPTON PATTERNS}

The present status of quark and lepton masses and couplings is summarized in
Figure 1. The top quark is the heaviest known, but the fractional error on its
mass, $m_t = 175 \pm 6~\G/c^2$ \cite{topmass}, is now the smallest for any
quark! A detailed pattern of charge-changing couplings among quarks occurs; in
addition to the dominant couplings $u \leftrightarrow d$, $c \leftrightarrow
s$, and $t \leftrightarrow b$, all the others are allowed, but with diminished
strengths, as shown in Table 1. 

\begin{figure} 
\centerline{\epsfysize = 2in \epsffile {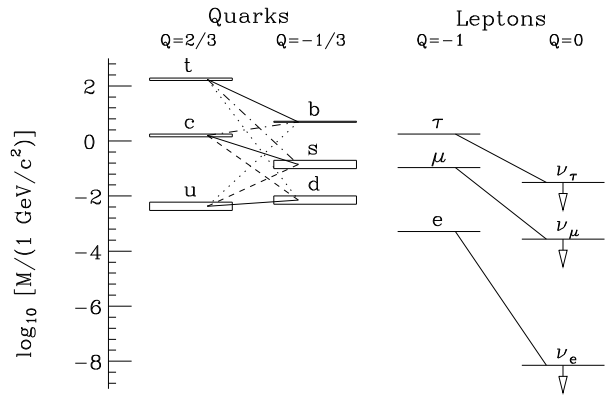}}
\caption{Patterns of charge-changing weak transitions among quarks and leptons.
Direct evidence for $\nu_\tau$ does not yet exist. The strongest inter-quark
transitions correspond to the solid lines, with dashed, dot-dashed, and dotted
lines corresponding to successively weaker transitions. Upper bounds on
neutrino masses are indicated by arrows.} 
\label{fig:qandl}
\end{figure}

\begin{table}
\caption{Relative strengths of charge-changing weak transitions.}
\begin{tabular}{c c l} \hline
Relative & Trans- & Source of information \\
ampl.    & ition  & ~~~~~~~~(example) \\ \hline
$\sim$ 1 & $u \leftrightarrow d$ & Nuclear $\beta$-decay \\
$\sim$ 1 & $c \leftrightarrow s$ & Charm decays \\
$\sim 0.22$ & $u \leftrightarrow s$ & Strange particle decays \\
$\sim 0.22$ & $c \leftrightarrow d$ & Neutrino charm prod. \\
$\sim 0.04$ & $c \leftrightarrow b$ & $b$ decays \\
$\sim 0.003$ & $u \leftrightarrow b$ & Charmless $b$ decays \\
$\sim$ 1 & $t \leftrightarrow b$ & Dominance of $t \to W b$ \\
$\sim 0.04$ & $t \leftrightarrow s$ & Only indirect evidence \\
$\sim 0.01$ & $t \leftrightarrow d$ & Only indirect evidence \\ \hline
\end{tabular}
\end{table}

The couplings in Table 1 are encoded in the unitary $3 \times 3$
Cabibbo-Kobayashi-Maskawa (CKM) matrix.  We now explore the current status of
its parameters.  More complete discussions may be found in \cite{AL},
\cite{MN}, and \cite{Cargese}. 

\section{CKM MATRIX AND PARAMETERS}

The CKM matrix $V$ describing the charge-changing weak transitions of
left-handed  quarks may be written as \cite{WP}
$$
V \approx \left[ \matrix{1 - \lambda^2/2 & \lambda & A \lambda^3 (\rho -
i \eta) \cr
- \lambda & 1 - \lambda^2 /2 & A \lambda^2 \cr
A \lambda^3 (1 - \rho - i \eta) & - A \lambda^2 & 1 \cr } \right].
$$
The quantity $\lambda = 0.2205 \pm 0.0018 = \sin \theta_c$ \cite{strange,PDG}
expresses the suppression of $s \to u$ decays with respect to $d \to u$ decays
\cite{Cab,GL}. This parameter is sufficient to describe the $u,~d,~s$, and $c$
couplings via the upper left $2 \times 2$ submatrix of $V$ \cite{charm}.

When a third family of quarks is added, three more parameters are needed.  One
may express them in terms of (1) the strength characterizing $b \to c$ decays,
$A \lambda^2 = 0.0393 \pm 0.0028$ \cite{AL,Gib} so that $A = 0.808 \pm 0.058$
(see \cite{MN}, \cite{ALK} for slightly different values); (2) the magnitude of
the $b \to u$ transition element measured in charmless $b$ decays,
$V_{ub}/V_{cb} = 0.08 \pm 0.016$ \cite{AL} (see also \cite{Kim}) so that 
\beq \label{eqn:Vubcon}
(\rho^2 + \eta^2)^{1/2} = 0.363 \pm 0.073~~;
\eeq
(3) the phase of $V_{ub} = A \lambda^3 (\rho - i \eta)$. The unitarity of the
CKM matrix implies that the scalar product of the complex conjugate of any row
with any other row should vanish, e.g., 
\beq \label{eqn:ur}
V_{ud}^* V_{td} + V_{us}^* V_{ts} + V_{ub}^* V_{tb} = 0 ~~~.
\eeq
Since $V_{ud}^* \approx 1,~V_{us}^* \approx \lambda,~V_{ts} \approx - A
\lambda^2$, and $V_{tb} \approx 1$ we have $V_{td} + V_{ub}^* = A \lambda^3$,
expressing the least-known CKM elements in terms of relatively well-known
parameters.  This result can be visualized as a triangle in the complex plane.
Dividing (\ref{eqn:ur}) by $A \lambda^3$, since $V_{ub}^*/ A \lambda^3 = \rho +
i \eta, ~~ V_{td} / A \lambda^3 = 1 - \rho - i \eta$, one obtains a triangle of
the form shown in Fig.~2. In this figure the angles $\alpha, \beta$, and
$\gamma$ are defined as in  \cite{NQ}.  Each can be measured using $B$ meson
decays. 

The value of $V_{ub}^* / A \lambda^3$ may then be depicted as a point in the
$(\rho,\eta)$ plane. The major remaining ambiguity in the determination of the
CKM matrix elements concerns the phase of $V_{ub}$, or the shape of the
unitarity triangle. The answer depends on the value of $V_{td}$. In order to
learn about this one must resort to indirect means, which involve loop
diagrams. 
 
\begin{figure}
\centerline{\epsfysize = 1.3in \epsffile {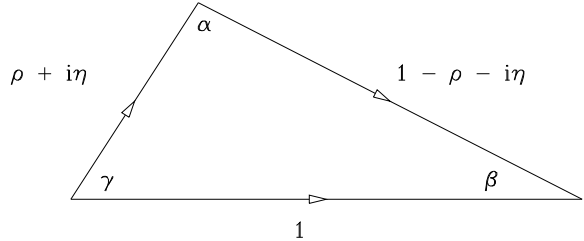}}
\caption{Unitarity triangle for CKM elements. We show in the complex plane the
relation (\protect\ref{eqn:ur}) divided by the normalizing factor $A
\lambda^3$.} 
\end{figure}

\subsection{$B^0 - \bar B^0$ mixing}

Box diagrams involving $u,~c,$ and $t$ in loops contribute to the virtual $b
\bar d \leftrightarrow d \bar b$ transitions which mix $\bar B^0$ and $B^0$.
The leading contribution at high internal momentum in these diagrams cancels as
a consequence of (\ref{eqn:ur}).  The remaining contribution is dominated by
the top quark since all products of CKM elements $V_{qb} V_{qd}^*$ are of order
$\lambda^3$ while $m_t \gg m_c,~m_u$.  The expression for the splitting between
mass eigenstates is then \cite{IL} 

\beq \label{eqn:dmd}
\Delta m = {G_F^2 \over 6 \pi^2} |V_{td}|^2 M_W^2 m_B f_B^2 B_B \eta_B S \left(
{m_t^2 \over M_W^2} \right),
\eeq
where
\beq \label{eqn:sdef}
S(x) \equiv {x \over 4} \left[ 1 + {3-9x \over (x-1)^2} + {6x^2\ln x \over
(x-1)^3} \right].
\eeq

We take $m_t = 175 \pm 6~\G/c^2$ \cite{topmass}, $M_W = 80.34 \pm 0.10~\G/c^2$
\cite{Cargese}, $m_B = 5.279~\G/c^2$ (see \cite{PDG}), and $f_B \sqrt{B_B} =
200 \pm 40$ MeV \cite{AL,MN}.  Here $f_B$ is the $B$ meson decay constant,
defined so that the matrix element of the weak axial-vector current $A_\mu
\equiv \bar b \gamma_\mu \gamma_5 d$ between a $B^0$ meson and the vacuum is
$\langle 0 | A_\mu | B^0(p) \rangle = i p_\mu f_B$.  With this normalization,
the decay constants of the light pseudoscalar mesons are $f_\pi = 131$ MeV and
$f_K = 160$ MeV.  The decay constants express the amplitude for the
corresponding quark and antiquark (e.g., $b$ and $\bar d$) to be found at a
point, as they must in order to participate in the short-distance process
associated with the box diagrams. 

The factor $B_B$ expresses the degree to which the box diagrams provide the
contribution to $B - \bar B$ mixing.  An estimate in lattice gauge theories
\cite{latBB} is $B_B = 1.16 \pm 0.08$.  Finally, $\eta_B = 0.55$ is a QCD
correction.  All quantities are quoted in the same consistent renormalization
scheme \cite{QCDB}. 

The first evidence for mixing of nonstrange $B$'s was obtained by the ARGUS
Collaboration \cite{ARmix}.  The large mixing amplitude, $\Delta m
/\Gamma \simeq 0.7$ (where $\Delta m$ is the mass difference between mass
eigenstates and $\Gamma$ is the $B$ meson decay rate), was one early indication
of a very heavy top quark. With the rapidly moving $B$ mesons and the fine
vertex information now available at LEP \cite{Gib,LEPmix}, CDF \cite{CDFmix},
and SLD \cite{SLDmix} (see also \cite{BH}), it has become possible to directly
observe time-dependent $B^0 - \bar B^0$ oscillations with a modulating factor
$\sin(\Delta m t)$ (where $t$ is the proper decay time). The current world
average \cite{Dmavg} is $\Delta m_d = 0.470 \pm 0.017~{\rm ps}^{-1}$, where the
subscript refers to the mixing between $B^0 \equiv \bar b d$ and $\bar B^0
\equiv b \bar d$.  Using (\ref{eqn:dmd}) and the parameters mentioned above, we
can then obtain an estimate of $|V_{td}|$, which leads, once we factor out a
term $A \lambda^3$, to the constraint 
\beq \label{eqn:Bmixcon}
|1 - \rho - i \eta | = 1.01 \pm 0.22~~~.
\eeq  
This result can be plotted in the $(\rho,\eta)$ plane as a band bounded by
circles with centers at (1,0).

Now that the top quark mass is known so precisely, the dominant source of error
in Eq.~(\ref{eqn:Bmixcon}) is uncertainty in the $B$ meson decay constant
$f_B$. Although the desired quantity $f_B$ has not yet been measured directly,
a number of results on the decay constant $f_{D_s}$ have appeared over the past
few years \cite{WA75,CLEOfDs,BESfDs,E653,L3fDs}; these serve to check
calculations of $f_B$.  The values of $f_{D_s}$ obtained by measuring the rates
for $D_s \to \mu \nu$ (and sometimes $\tau \nu$) are summarized in Table 2. The
errors are statistical, systematic, and (where shown) branching ratio of
calibrating mode.  These are added in quadrature to obtain the value of
$\sigma$ in the last column. One can also estimate $f_{D_s}$ by assuming
factorization in certain $B$ decays in which the charged weak current produces
a $D_s$ \cite{Cargese}. The results are consistent with the above average.

\begin{table}
\caption{Values of $f_{D_s}$ obtained by measuring the rates for $D_s \to \mu
\nu$ and/or $\tau \nu$}
\begin{tabular}{c c c} \hline
Expt.~(Ref.) & $f_{D_s}$ (MeV) & $\sigma$ (MeV) \\ \hline
WA75 \protect \cite{WA75} &   $232 \pm 45 \pm 20 \pm 48$ & 69 \\
CLEO \protect \cite {CLEOfDs} & $284 \pm 30 \pm 30 \pm 16$ & 45 \\
BES \protect \cite{BESfDs} & $430^{+150}_{-130} \pm 40$ & 146 \\
E653 \protect \cite{E653} & $194 \pm 35 \pm 20 \pm 14$ & 43 \\
L3 \protect \cite{L3fDs} & $311 \pm 58 \pm 32 \pm 16$ & 68 \\ \hline
Average & 253 & 26 \\ \hline
\end{tabular}
\end{table}

One $D \to \mu \nu$ event has been detected by the BES group, leading to $f_D =
300^{+180+80}_{-150-40}$ MeV \cite{BESfD}.  This result is consistent with the
previous upper bound of 290 MeV (90\% c.l.) obtained by Mark III \cite{MkIII}. 
Flavor SU(3)-breaking estimates in lattice gauge theories \cite{FBL} and quark
models \cite{FBQ} imply $f_D/f_{D_s} = 0.8$ to 0.9. 

Lattice gauge theories and quark models are both are converging on the range
$f_B \simeq 180 \pm 40$ MeV, implying branching ratios $B(B \to \tau \nu)
\simeq (1/2) \times 10^{-4}$ and $B(B \to \mu \nu) \simeq 2 \times 10^{-7}$
\cite{JRCP}.  These small rates pose a challenge to $B$ factories. The same
SU(3) estimates for the ratio of $f_D/f_{D_s}$ also give a very similar ratio
of $f_B/f_{B_s}$.  Equality of these two ratios is expected within a few
percent \cite{Grin}. 

\subsection{CP-violating $K^0 - \bar K^0$ mixing}

The $\k$ and $\bk$ are strong-interaction eigenstates of opposite strangeness.
However, since the weak interactions do not conserve strangeness, they pick out
linear combinations of $\k$ and $\bk$ in decay processes \cite{GP}.  As of
1957, when the weak interactions were understood to violate charge-conjugation
invariance C and spatial reflection P but to preserve their product CP, one
expected \cite{KCP} the linear combination $K_1^0 \equiv (\k + \bk)/\s$, with
even CP, to have a much more rapid decay rate since it could decay to the
CP-even final state of two pions.  The orthogonal linear combination $K_2^0
\equiv (\k - \bk)/\s$, with odd CP (seen in 1956 \cite {KL}), would live much
longer since it was forbidden by CP invariance to decay to two pions and would
have to decay to three pions or a pion and a lepton-neutrino pair. 

In 1964 J. Christenson, J. Cronin, V. Fitch, and R. Turlay reported that in
fact the long-lived neutral kaon {\it did} decay to two pions, with an
amplitude whose magnitude is about $2 \times 10^{-3}$ that for the short-lived
$K \to 2 \pi$ decay \cite{CCFT}. One then can parametrize the mass eigenstates
as 
\beq
K_S \simeq K_1 + \epsilon K_2~,~~
K_L \simeq K_2 + \epsilon K_1~~~,
\eeq
where $|\epsilon| \simeq 2 \times 10^{-3}$ and the phase of $\epsilon$ turns
out to be about $\pi/4$.  The parameter $\epsilon$ encodes all current
knowledge about CP violation in the neutral kaon system.  Where does it come
from? 

One possibility, proposed \cite{sw} immediately after the discovery and still
not excluded, is a ``superweak'' CP-violating interaction which directly mixes
$\k = d \bar s$ and $\bk = s \bar d$.  This interaction would have no other
observable consequences since the $\k - \bk$ system is so sensitive to it!

The presence of three quark families \cite{KM} poses another opportunity for
explaining CP violation through box diagrams involving $u,~c$, and $t$ quarks.
With three quark families, phases in complex coupling coefficients cannot be
removed by redefinition of quark phases.  Within some approximations
\cite{JRCP}, the parameter $\epsilon$ is directly proportional to the imaginary
part of the mixing amplitude.  Its magnitude (see \cite{CL} or \cite{TASI} for
a calculation in the limit of $m_t \ll M_W$) is \cite{IL} 
$$
|\epsilon| \simeq \frac{G_F^2 m_K f_K^2 B_K M_W^2}{\s (12 \pi^2)
\Delta m_K} \times
$$
\beq \label{eqn:eps}
[\eta_1 S(x_c) I_{cc} + \eta_2 S(x_t) I_{tt} + 2 \eta_3 S(x_c,x_t)
I_{ct}]~,
\eeq
where $I_{ij} \equiv {\rm Im}(V_{id}^* V_{is} V_{jd}^* V_{js})$.
In order to evaluate these expressions we need to work to sufficiently high
order in small parameters in $V$. The application of the unitarity relation to
the first and second rows tells us that $V_{cd} = - \lambda - A^2 \lambda^5
(\rho + i \eta).$  We then find $I_{cc} = - 2 A^2 \lambda^6 \eta$, $I_{ct} =
A^2 \lambda^6 \eta$, and $I_{tt} = 2 A^2 \lambda^6 \eta [A^2 \lambda^4 (1-
\rho)]$. The factors $\eta_1 = 1.38,~\eta_2=0.57,~\eta_3=0.47$ are QCD
corrections \cite{QCDK}, while $x_i \equiv m_i^2/M_W^2.$  The function $S(x)$
was defined in Eq.~(\ref{eqn:sdef}), while 
\beq 
S(x,y) \equiv xy \left\{ 
\begin{array}{c} \left[ {1 \over 4} + {3 \over 2(1-y)} - {3 \over
4(1-y)^2} \right] {\ln y \over y-x} \\
+ (y \leftrightarrow x) - {3 \over 4(1-x)(1-y)}
\end{array} \right\}.
\eeq

Eq.~(\ref{eqn:eps}) may then be rewritten (cf.~\cite{HRS}) as
$$
|\epsilon| = 4.39 A^2 B_K \eta ~ \times
$$
\beq
[\eta_3 S(x_c,x_t) - \eta_1 S(x_c) + \eta_2
A^2 \lambda^4 (1-\rho) S(x_t) ]~.
\eeq
Using the experimental values \cite{PDG} $|\epsilon| = (2.28 \pm 0.02) \times
10^{-3}$, $f_K = 160~\M$, $\Delta m_K = 3.49 \times 10^{-15}~\G$, and $m_K =
0.4977~\G$, the value $B_K = 0.75 \pm 0.15$ \cite{BKlat}, and the top quark mass
$\bar m_t (M_W) = 165 \pm 6$ GeV/$c^2$ appropriate for the loop calculation
\cite{AL}, we find that CP-violating $K - \bar K$ mixing leads to the
constraint 
\beq \label{eqn:Kmixcon}
\eta(1 - \rho + 0.44) = 0.51 \pm 0.18~~~,
\eeq
where the term $1 - \rho$ corresponds to the loop diagram with two top quarks,
and the term 0.44 corresponds to the additional contribution of charmed quarks.
 The major source of error on the right-hand side is the uncertainty in the
parameter $A \equiv V_{cb}/\lambda^2$. Eq.~(\ref{eqn:Kmixcon}) can be plotted
in the $(\rho,\eta)$ plane as a band bounded by hyperbolae with foci at
(1.44,0). 

The constraints (\ref{eqn:Vubcon}), (\ref{eqn:Bmixcon}), and
(\ref{eqn:Kmixcon}) define the allowed region of parameters shown in Fig.~3.
The boundaries shown are $1 \sigma$ errors, but are dominated by theoretical
uncertainties in each case.

\begin{figure}
\centerline{\epsfysize = 1.6in \epsffile {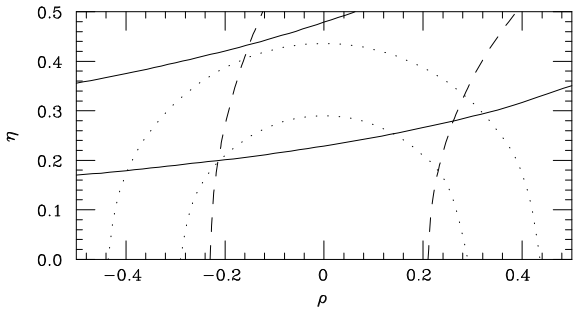}}
\caption{Region in the $(\rho,\eta)$ plane allowed by constraints on
$|V_{ub}/V_{cb}|$ (dotted semicircles), $B^0 - \bar B^0$ mixing (dashed
semicircles), and CP-violating $K - \bar K$ mixing (solid hyperbolae).}
\end{figure}
 
A large region centered about $\rho \simeq 0$, $\eta \simeq 0.35$ is permitted.
Nonetheless, the CP violation seen in kaons could be due to an entirely
different source, such as a superweak mixing of $K^0$ and $\bar K^0$ \cite{sw}.
In that case one could probably still accommodate $\eta = 0$, and hence a real
CKM matrix, by going slightly outside the $1 \sigma$ bounds based on
$|V_{ub}/V_{cb}|$ or $B - \bar B$ mixing.  In order to confirm the predicted
nonzero value of $\eta$, we turn to other experimental possibilities. Many of
these, such as the search for direct CP violation in $K^0 \to \pi \pi$ decays
and the search for rare kaon decays, are covered elsewhere in this Workshop
\cite{Ciuchini,Buchalla}; here we concentrate on $B$ decays. 

\section{CKM TESTS WITH $B$ MESON DECAYS}

A number of experimental facilities can or will be able to make incisive tests
of the CKM picture by studying hadrons containing the $b$ quark. LEP has
finished productive years of running on the $Z^0$, in which millions of $b \bar
b$ pairs were produced. SLD has the advantage of electron polarizability which
partially compensates for a lower luminosity than LEP. CESR at Cornell
continues to set luminosity records and is aiming for ${\cal L} > 10^{33}$
cm$^{-2}$ s$^{-1}$ at the $\Upsilon(4S)$, a copious source of $B \bar B$ pairs.
The collider detectors at Fermilab, D0 and particularly CDF, have shown the
utilitiy of $B$ studies in 1.8 TeV $\bar p p$ collisions. Under construction
are experiments at HERA-B, PEP-II, and KEK-B, the first to study fixed-target
$b$ production by 800 GeV protons and the latter two to study asymmetric $e^+
e^-$ collisions at the $\Upsilon(4S)$.  In the farther future lie projects at
LHC-B and possibly Fermilab. 

We have already mentioned the importance of $B^0 - \bar B^0$ mixing as a
validation of the CKM description of couplings. In this section we discuss some
other aspects of $B$ decays crucial to testing the CKM picture \cite{BCP}. 

Decays to CP eigenstates such as $B^0 \to J/\psi K_S$ form the core of the
program for discovering CP violation in the $B$ system.  A key feature of such
studies is the ability to distinguish an initially produced $B^0$ from a $\bar
B^0$.  Progress has been made by the CDF Collaboration in its study of methods
for ``tagging'' the flavor of a produced $B$ \cite{Schmidt}. 

The strange $B$ mesons $B_s \equiv \bar b s$ and $\bar B_s \equiv b \bar s$ are
expected to mix with one another with a large amplitude, such that $\Delta
m/\Gamma \gg 1$.  The mass eigenstates are expected to be approximate
eigenstates of CP:  $CP~B_s^{(\pm)} = \pm 1$.  We shall discuss one method
\cite{DDLR} for separating such eigenstates on the basis of angular
distributions in $B_s^{(\pm)} \to J/\psi \phi \to e^+ e^- K^+ K^-$. If $\Delta
m$ is large for strange $B$'s, so is $\Delta \Gamma$, with the CP-even state
expected to have a 10 or 20\% more rapid decay rate than the CP-odd one
\cite{Beneke}. 

Some progress in understanding the role of ``penguin'' diagrams, which give
rise to induced $b \to d$ and $b \to s$ transitions, has been made in recent
years.  We shall discuss penguin amplitudes with particular reference to their
role in decays of $B$ mesons to pairs of light mesons, such as $B \to \pi \pi$,
$\pi K$, $\eta K$, etc.  From these decays it is possible to learn about phases
of CKM elements. 

\subsection{Decays to CP eigenstates}

By comparing rates of decays to CP eigenstates for a state produced as $B^0$
and one produced as a $\bar B^0$, one can directly measure angles in the
unitarity triangle of Fig.~2.  Because of the interference between direct
decays (e.g., $B^0 \to J/\psi K_S$) and those which proceed via mixing (e.g.,
$B^0 \to \bar B^0 \to J/\psi K_S$), these processes are described by
time-dependent functions whose difference when integrated over all time is
responsible for the rate asymmetry. Thus, if we define 
\beq
C_f \equiv \frac{\Gamma(B_{t=0} \to f) - \Gamma(\bar B_{t=0} \to f)}
{\Gamma(B_{t=0} \to f) + \Gamma(\bar B_{t=0} \to f)}~,
\eeq
we have, in the limit of a single direct contribution to decay amplitudes,
\beq \label{eqn:asymms}
A(J/\psi K_S, \pi^+ \pi^-) = - \frac{x_d}{1+x_d^2} \sin(2\beta,2\alpha)~,
\eeq
where $x_d \equiv \Delta m(B^0)/\Gamma(B^0)$.  This limit is expected to be
very good for $J/\psi K_S$, but some correction for penguin contributions (to
be discussed below) is probably needed for $\pi^+ \pi^-$.

To see this behavior in more detail \cite{Isi}, we note that the time-dependent
partial rates for a state which is initially $B^0~(\bar B^0)$ to decay to a
final state $f$ may be written as 
$$
d \Gamma [B^0 (\bar B^0) \to f]/dt  \sim e^{- \Gamma t}
[1 \mp {\rm Im} \lambda_0 \sin (\Delta m t)]~~~,
$$
where we have neglected $\Delta \Gamma / \Gamma$ in comparison with $\Delta m /
\Gamma$. This step is justified for $B$'s, though not for $K$'s.  The final
states to which both $B^0$ and $\bar B^0$ can decay are only a small fraction
of those to which $B^0$ or $\bar B^0$ normally decay, so one should expect
similar lifetimes for the two mass eigenstates.  Integration of this equation
gives 
\beq \label{eqn:asy}
C_f = \frac{-x_d}{1 + x_d^2} {\rm Im}\lambda_0(f)
\eeq
for the total asymmetry.  For the final states mentioned, $\lambda_0(J/\psi
K_S) = - e^{- 2 i \beta}$ and $\lambda_0(\pi^+ \pi^-) = e^{2 i \alpha}$.  The
extra minus sign in the first relation is due to the odd CP of the $J/\psi K_S$
final state. 

The asymmetry (\ref{eqn:asy}) is suppressed both when $\Delta m/\Gamma$ is very
small and when it is very large (e.g., as is expected for $B_s$). For $B_s$, in
order to see an asymmetry, one must not integrate with respect to time.
Experiments planned with detection of $B_s$ as their focus will require precise
vertex detection to measure mixing as a function of proper time.  For $B^0$, on
the other hand, the value of $x/(1+x^2)$ for $x=0.7$ is 0.47, very close to its
maximum possible value of 1/2 for $x=1$. 

When more than one eigenchannel contributes to a decay, terms of the form $\cos
(\Delta m t)$ as well as $\sin (\Delta m t)$ can appear \cite{PP}. These
complicate the analysis, but information can be obtained from them \cite{pipi}
on the relative contributions of various channels to decays. 

\subsection{Neutral $B$ flavor tagging}

In searching for rate asymmetries in decays to CP eigenstates of neutral $B$'s
one must know whether they were $B^0$ or $\bar B^0$ at the time of production,
since the final state does not tell us this.  Several methods are available for
``tagging'' the flavor of the produced $B$. 

1.  In uncorrelated $b \bar b$ production, as occurs in hadronic or high-energy
$e^+ e^-$ collisions, one can identify the flavor of a neutral meson (e.g., a
$\bar B^0$ meson containing a $b$ quark) by means of the flavor of the hadrons
produced in association: in this case $B^0$, $B^+$, $B_s$, $\bar \Lambda_b$,
etc.  The semileptonic decays of the quarks in these hadrons will lead to a
lepton (typically only $e$ or $\mu$ are useful) whose sign ``tags'' the flavor
of the opposite-side hadron.  Charged kaons may also have some utility in
flavor tagging, through the chain $b \to c \to s$ \cite{Bellini}.

2.  Just above threshold in $e^+ e^-$ annihilations, a $B^0 \bar B^0$ pair
is produced in a state of odd $C$, so the decay products of the
$B^0$ and $\bar B^0$ are highly correlated.  If the initial $B^0$ decays at
time $t$ and the initial $\bar B^0$ decays at $\bar t$, the decay
asymmetry depends on $\sin \Delta m(t - \bar t)$, and
hence is odd in $t - \bar t$.  When integrated over $t$ and $\bar t$, the
asymmetry thus vanishes!  Consequently, one needs to measure the
time-dependence of the individual decays, or at least to be sensitive to the
sign of the time difference.  This requirement has spawned the asymmetric
``B factories'' now under construction at SLAC and KEK, where the Lorentz
boost of the center-of-mass spreads out the decays so they can be resolved from
one another.  At the symmetric CESR machine, the necessary spatial resolution
may be achievable despite the short decay path of the neutral $B$'s (only 30
$\mu$m!) using silicon vertex detectors and very flat beams \cite{KB}. 

3.  A third method for tagging the flavor of neutral $B$'s at the time of
production is to use the fact that a charged pion of a given sign is most
likely to be associated with a given flavor of neutral $B$ \cite{tags}. Thus,
when a $b$ quark fragments into a $\bar B^0$ containing a $\bar d$ quark, a $d$
quark is available nearby in phase space to be incorporated into a $\pi^-$.  As
a result, one will expect a $\bar B^0$ to be accompanied more often by a
$\pi^-$ than by a $\pi^+$.  This same correlation is expected in resonance
production:  $\bar B^0$ can resonate with $\pi^-$ to form a non-exotic ($q \bar
q$) state $b \bar u$, but not with a $\pi^+$.  A similar argument favors $B^0
\pi^+$ over $B^0 \pi^-$ combinations. 

Several LEP groups have seen the expected correlations \cite{LEPtags}.  The CDF
Collaboration has now used this correlation to provide an independent
measurement of $\Delta m$ in $B^0 - \bar B^0$ oscillations \cite{Schmidt}.
Together with leptonic flavor tagging (the method mentioned in paragraph 1
above), this method shows promise for identifying the flavor of at least
several percent of all hadronically produced neutral $B$'s, thus opening the
possibility for observing a CP-violating rate asymmetry in $B^0$ or $\bar B^0
\to J/\psi K_S$ in the next CDF run. 

\subsection{Mixing of strange $B$'s}

The mixing between strange $B$'s due to box diagrams is
considerably enhanced relative to that between nonstrange $B$'s: 
\beq
\frac{\Delta m_s}{\Delta m_d} = \frac{f_{B_s}^2 B_{B_s}}{f_B^2 B_B}
\left| \frac{V_{ts}}{V_{td}} \right|^2 \simeq 17 - 52~~~,
\eeq
where we have taken the expected ranges of decay constant and CKM element
ratios, and $\Delta m_s$ refers to mixing between the $B_s \equiv \bar b s$
and $\bar B_s \equiv b \bar s$.  Alternatively, we may retrace the evaluation
of $\Delta m$ for nonstrange $B$'s, replacing appropriate quantities in
Eq.~(\ref{eqn:dmd}), to derive an analogous expression for $\Delta m_s$, which
we then evaluate directly.  For $V_{ts} = 0.040 \pm 0.004$, $m_{B_s} =
5.37~\G/c^2$, $f_{B_s} \sqrt{B_{B_s}} = 225$ MeV, $\eta_{B_s} = 0.6 \pm 0.1$,
and \cite{BH} $\tau_{B_s} \equiv 1/\Gamma_s = 1.55 \pm 0.10$ ps, we find
$\Delta m_s/\Gamma_s = 22 \pm 6$, with an additional 40\% error associated with
$f_{B_s}^2 B_{B_s}$.  This result implies many particle-antiparticle
oscillations in a decay lifetime, requiring good vertex resolution and highly
time-dilated $B_s$'s for a measurement.  The present experimental bound $\Delta
m_s > 9.2$ ps$^{-1}$ based on combining ALEPH and DELPHI results \cite{dms}
begins to restrict the parameter space in an interesting manner. 

The large value of $\Delta m_s$ entails a value of $\Delta \Gamma_s$ between
mass eigenstates of strange $B$'s which may be detectable.  After all, the
short-lived and long-lived neutral kaons differ in lifetime by a factor of 600.
Strong interactions and the presence of key channels (e.g., $\pi \pi$) are a
crucial effect in strange particle (e.g., $\k$ and $\bk$) decays.  While the
$b$ quark decays as if it is almost free, so that strong interactions are
much less important, a corresponding difference in lifetimes
for strange $B$'s of the order of $10 - 20\%$ is not unlikely \cite{Blifes}. 

In the ratio $\Delta m_s/\Delta \Gamma_s$, uncertainties associated with the
meson decay constants cancel, and in lowest order (before QCD corrections) one
finds \cite{IsiBP} $\Delta m_s/\Delta \Gamma_s \simeq {\cal O}(-[1/\pi]
[m_t^2/m_b^2]) \simeq - 200$.  The heavier state is expected to be the
longer-lived one, as in the neutral kaon system. The top quark does not
contribute to the width difference associated with the imaginary part of the
box graphs, since no $t \bar t$ pairs are produced in $B_s$ decays. 

Aside from small CP-violating effects, the mass eigenstates of strange $B$'s
correspond to those $B_s^{(\pm)}$ of even and odd CP. The decay of a $\bar B_s$
meson via the quark subprocess $b (\bar s) \to c \bar c s (\bar s)$ gives rise
to predominantly CP-even final states \cite{CPeven}, so the CP-even eigenstate
should have a greater decay rate. One calculation \cite{Blifes} gives 
\beq \label{eqn:widthdiff}
\frac{\Gamma(B_s^{(+)}) - \Gamma(B_s^{(-)}) }{\overline \Gamma} \simeq 0.18
\frac{f_{B_s}^2}{(200~\M)^2}~~~,
\eeq
while a more recent estimate \cite{Beneke} is $0.16^{+0.11}_{-0.09}$.
The lifetime difference between CP-even and CP-odd strange $B$'s thus can
provide useful information on $f_{B_s}$, and hence indirectly on the weak
interactions at short distances.

\subsection{Isolating strange $B$ eigenstates}

One way to separate strange-$B$ CP eigenstates from one another \cite{DDLR} is
to study angular distributions in $B_s \to J/\psi + \phi \to e^+ e^- K^+ K^-$
(or $\mu^+ \mu^- K^+ K^-$).  The $J/\psi$ and $\phi$ are both spin-1 particles
and hence can be produced in states of orbital angular momenta $L = 0,~1$, and
2 from the spinless $B_s$ decay.  Suitably normalized $L = 0,~1$, and 2
amplitudes $S$, $P$, $D$ can be defined such that $|S|^2 + |P|^2 + |D|^2 = 1$.
$L=1$ corresponds to $P = CP = -$, while $L = 0$ or 2 corresponds to $P = CP =
+$.  A simple transversity analysis \cite{Tr} permits one to separate the two
cases. 

In the $J/\psi$ rest frame, let the $x$ axis be defined by the direction of the
$\phi$, the $x-y$ plane be defined by the kaons which are its decay products,
and the $z$ axis be the normal to that plane.  Let the $e^+$ (or $\mu^+$) make
an angle $\theta$ with the $z$ axis.  Then CP-even final states give rise to an
angular distribution $1 + \cos^2 \theta$, while the CP-odd state gives rise to
$\sin^2 \theta$.  When both CP eigenstates are present in $B_s \to J/\psi
\phi$, one will see a gradual increase of the $\sin^2 \theta$ component
relative to the $1 + \cos^2 \theta$ component. More likely (if predictions
\cite{CPeven} are correct), the CP-even state will dominate, so that one will
measure mainly the lifetime of this eigenstate when following the
time-dependence of the decay.  The average decay rate $\bar \Gamma \equiv
(\Gamma_+ + \Gamma_-)/2$ (the subscripts denote CP eigenvalues) is measured in
flavor-tagged decays of $B_s = \bar b s \to \bar c + \ldots$ or $\bar B_s = b
\bar s \to c + \ldots$. 

A recent analysis \cite{CLEOTr} of $B \to J/\psi K^*$ has bearing on the $B_s
\to J/\psi \phi$ partial-wave structure.  The two processes are related by
flavor SU(3), involving a substitution $s \leftrightarrow d$ of the spectator
quark.  Thus, one expects the same partial waves in the two decays.  The CLEO
Collaboration has studied 146 $B \to J/\psi K^*$ decays in $3.36 \times 10^6~B
\bar B$ pairs produced at the Cornell Electron Storage Ring (CESR). The results
for this process are $|P|^2 = 0.21 \pm 0.14$ from a fit to the transversity
angle, and $|P|^2 = 0.16 \pm 0.08 \pm 0.04$ from a fit to the full angular
distribution.  This implies [via flavor SU(3)] that $B_s \to J/\psi \phi$ is
dominated by the CP-even final state, and thus the lifetime in this state
measures approximately $\tau(B_s^{(+)})$.  If any evidence for non-zero $|P|^2$
can be gathered in $B \to J/\psi K^*$, then $B_s \to J/\psi \phi$ should
exhibit the time-variation in the transversity-angle distribution mentioned
above \cite{DDLR}. 

\subsection{Processes dominated by penguin diagrams}

Although the unitarity of the CKM matrix implies flavor conservation for
charge-preserving electroweak interactions in lowest order, we have seen
that loop diagrams can induce flavor-changing charge-preserving interactions
in higher order.  Another example of this phenomenon is provided by the
``penguin'' diagram \cite{pen} illustrated in Fig.~4.  Although the penguin's
``leg'' is a gluon in this illustration, it can also be a photon or $Z$. When
the external quarks $x$ and $y$ have charge $-1/3$, the intermediate quarks
have charge 2/3 and can include the top quark.  Because of the top quark's
large mass, such penguin diagrams can be very important. 
 
\begin{figure}
\centerline{\epsfysize = 2 in \epsffile {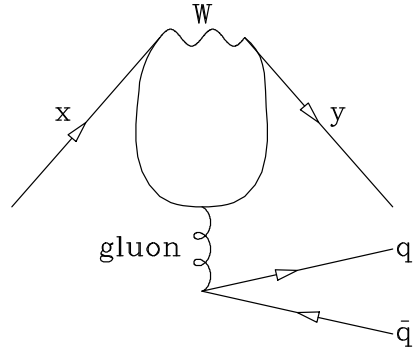}}
\caption{``Penguin'' diagram describing transition of a quark $x$ to another
quark $y$ with the same charge.  The intermediate quarks have charge
differing from $Q(x) = Q(y)$ by one unit.  Here $q = (u,d,s)$.}
\end{figure}

An example of a predicted penguin effect in $s \to d$ transitions is a phase
arising in the decays of neutral kaons to $\pi \pi$.  This phase can lead to a
``direct'' contribution to the ratios for CP-violating and CP-conserving
decays, in addition to that provided by the mixing parameter $\epsilon$
measured earlier. 

One may define (see \cite{Ciuchini} for details)
$$
\eta_{+-} \equiv \frac{A(K_L \to \pipe)}{A(K_S \to \pipe)};~
\eta_{00} \equiv \frac{A(K_L \to \poop)}{A(K_S \to \poop)}~;
$$
the effect of ``direct'' decays then shows up in a parameter $\epsilon'$
which causes $\eta_{+-}$ and $\eta_{00}$ to differ from one another:
\beq
\eta_{+-} = \epsilon + \epsilon'~~~;~~~\eta_{00} = \epsilon - 2 \epsilon'~~~.
\eeq
Since $\epsilon'$ and $\epsilon$ are expected to have approximately the same
phase (see, e.g., \cite{JRCP}), one expects
$| \eta_{+-} | \simeq | \epsilon | [ 1 + {\rm Re} (\epsilon '/\epsilon) ]$,
$| \eta_{00} | \simeq | \epsilon | [ 1 - 2 {\rm Re} (\epsilon'/\epsilon ]$,
and hence
$$
\frac{\Gamma (K_L \to 2 \pi^0)}{\Gamma ( K_S \to 2 \pi^0 )} /
\frac{\Gamma (K_L \to \pipe )}{\Gamma (K_S \to \pipe )} = 1 - 6~{\rm Re}
\frac{\epsilon '}{\epsilon} ~.
$$
Present expectations \cite{Ciuchini,epspth} are that $\epsilon'/\epsilon$ could
be a few parts in $10^4$ (but in any case $\epsilon'/\epsilon \le 10^{-3}$),
requiring the above ratio of ratios to be measured to about one part in $10^3$.
 Experiments now in progress at Fermilab and CERN should have the required
sensitivity.  The previous results of these experiments are: 
\beq
{\rm E731}~\cite{E731}: ~{\rm Re} (\epsilon '/ \epsilon) = (7.4 \pm 6.0)
\times 10^{-4} ~,
\eeq
\beq
{\rm NA31}~\cite{NA31}: ~{\rm Re} (\epsilon '/\epsilon) = (23.0 \pm 6.5)
\times 10^{-4} ~.
\eeq
Because of the cancelling effects of gluonic and ``electroweak'' penguins
(in which the curly line in Fig.~4 is a photon or $Z$), the actual magnitude
of $\epsilon'/\epsilon$ is difficult to estimate, so that one's best hope is
for a non-zero value within the rather large theoretical range, thereby
disproving the superweak model \cite{sw} of CP violation.

The contribution of the (gluonic) penguin diagram to the effective weak
Hamiltonian may be written (for $x,y$ equal to quarks of charge $-1/3$)
$$
{\cal H}_W^{\rm penguin} \simeq \frac{G_F}{\s}\frac{\alpha_s}{6 \pi} \left[
\xi_c \ln \frac{m_t^2}{m_u^2} + \xi_t \ln \frac{m_t^2}{m_u^2} \right] \times
$$
$$
\left[ (\bar y_L \gamma^\mu \lambda^a x_L)(\bar u \gamma_\mu \lambda^a u
+ \bar d \gamma_\mu \lambda^a d + \ldots) + \Hc \right]~,
$$
where $\xi_i \equiv V_{ix} V^*_{iy}$, and $\lambda^a$ are color SU(3) matrices
[$a = (1,\ldots,8)$] normalized so that Tr$(\lambda^a \lambda^b) = 2 
\delta^{ab}$.  The top quark is dominant in the flavor-changing processes
$b \to d$ and $b \to s$ (with corrections due to charm which can be important
in some cases \cite{BuFl}), while the charmed quark dominates the $s \to d$
process.  (The top quark plays a key role, however, in the electroweak penguin
contribution to this process \cite{epspth}.)  One may imitate the effect of
an infrared cutoff for the gluonic penguin graph by using a constituent-quark
mass $m_u \sim 0.3~\G/c^2$.

One can estimate the effect of the $b \to s q \bar q$ penguin graph; one finds
it is comparable to that of the $b \to u d \bar u$ ``tree'' contribution
$(G_F/\s)V_{ub}V^*_{ud} [\bar u \gamma_\mu (1 - \gamma_5) b][\bar d \gamma^\mu
(1 - \gamma_5) u]$.  The $b \to d q \bar q$ penguin contribution and the $b \to
u s \bar u$ tree contribution are both expected to be suppressed by
approximately one power of the Wolfenstein parameter $\lambda \sim 0.2$, as one
can see by comparing CKM elements. 

\subsection{Decays of $B$ mesons to pairs of light mesons}

One testing ground for the magnitude of penguin contributions occurs in the
decays of $B$ mesons to pairs of light mesons.    Thus, $B^0 \to \pi^+
\pi^-$ is expected to be dominated by the tree amplitude, $B^0 \to K^+ \pi^-$
is expected to be dominated by the penguin amplitude, and the rates of the two
processes should be similar.

In Fig.~5 we show a contour plot of the significance of detection by the CLEO
Collaboration \cite{Battle} of the decays $B^0 \to \pipe$ and $B^0 \to K^+
\pi^-$.  Evidence exists for a combination of $B^0 \to K^+ \pi^-$ and $\pi^+
\pi^-$ decays, generically known as $B^0 \to h^+ \pi^-$. The most recent
published result \cite{Wurt} is $B(B^0 \to h^+ \pi^-) = (1.8^{~+0.6~+0.2}
_{~-0.5~-0.3} \pm 0.2) \times 10^{-5}$. Although one still cannot conclude that
either decay mode is nonzero at the $3 \sigma$ level, the most likely solution
is roughly equal branching ratios (i.e., about $10^{-5}$) for each mode.  Only
upper limits exist for other modes of two pseudoscalars \cite{CLEOAPS}, but
these are consistent with predictions \cite{BPP}. 

\begin{figure}
\centerline{\epsfysize = 2.2 in \epsffile {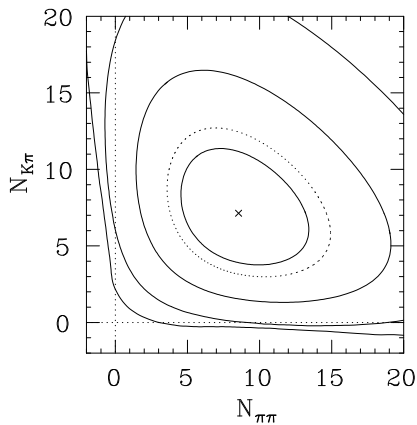}}
\caption{Significance of detection of $B^0 \to \pipe$ and $B^0 \to K^+ \pi^-$
by the CLEO Collaboration.  The solid curves are the $n \sigma$ contours and
the dotted curve is the $1.28 \sigma$ contour.} 
\end{figure}

Penguin diagrams play a number of roles in $B$ decays \cite{JRCP}. We enumerate
several of them. 

1.  The process $B^+ \to K^0 \pi^+$ is expected to be almost completely due to
the penguin graph. By comparison with $B^0 \to K^+ \pi^-$, where the penguin
graph is expected to be the main contribution, one expects $B(B^+ \to K^0
\pi^+) \simeq 10^{-5}$.  The weak phase of the process (which changes sign
under charge-conjugation) thus is expected to be Arg($V_{tb}^* V_{ts}) = \pi$,
so that the charge-conjugate process has the same weak phase. As a result, one
can separate strong final-state interaction phases from weak phases and obtain
estimates of quantities like the angle $\gamma = {\rm Arg}(V_{ub}^*)$ in Fig.~2
by comparing rates for $B^+ \to (K^0 \pi^+,~K^+ \pi^0,~K^+ \eta,~K^+ \eta')$
with the corresponding $B^-$ rates \cite{GReta}. One can also obtain this
information by measuring the time-dependence in $B^0 (\bar B^0) \to \pipe$ and
the rates for $B^0 \to K^+ \pi^-$, $B^+ \to K^0 \pi^+$, and the
charge-conjugate processes \cite{DGR}.  The weak phases of the major amplitudes
contributing to these decays are summarized in Table 3.  The relative weak
phase of tree and penguin amplitudes for strangeness-preserving decays is
$\gamma + \beta = \pi - \alpha$ (assuming the unitarity triangle to be valid),
while the corresponding relative phase for strangeness-changing decays (aside
from a sign) is just $\gamma$.  As a result, one can measure both $\alpha$ and
$\gamma$. 

\begin{table}
\caption{Phases of amplitudes contributing to decays of $B$ mesons to
$\pi \pi$ and $K \pi$.  Here $\Delta S$ refers to the change of strangeness
in the process.}
\begin{center}
\begin{tabular}{c|c c|c c} \hline
  & \multicolumn{2}{c|}{``Tree''} & \multicolumn{2}{c}{``Penguin''} \\
\cline{2-3} \cline{4-5}
 $|\Delta S|$ & CKM             &   Phase  & CKM               & Phase \\
              & elements        &          & elements          & \\ \hline
       0     & $V^*_{ub}V_{ud}$ & $\gamma$ & $V^*_{tb} V_{td}$ & $-\beta$ \\
       1     & $V^*_{ub}V_{us}$ & $\gamma$ & $V^*_{tb} V_{ts}$ & $\pi$ \\
\hline 
\end{tabular}
\end{center}
\end{table}

2.  A number of processes (in addition to the decay $B^+ \to K^0 \pi^+$
mentioned above) are dominated by penguin graphs.  By comparing the rates for
strangeness-preserving and strangeness-changing processes, one can measure the
ratio $|V_{td}/V_{ts}|$ \cite{GRP}.  Examples of useful ratios are $B(B^+ \to
\bar K^{*0} K^+)/B(B^+ \to \phi K^+)$ and $B(B^+ \to \bar K^{*0} K^{*+})/B(B^+
\to \phi K^{*+})$. 

3.  We mentioned that the time-integrated rate asymmetry in $B \to
\pipe$ could provide information on the angle $\alpha$ of the unitarity
triangle.  The most direct test is based on the assumption that the tree
process $\bar b \to \bar u u \bar d$ is the only direct contribution to this
decay.  However, ``penguin pollution'' \cite{PP} (due to the $b \to d$
transition) makes the analysis less straightforward, even thought the penguin
amplitude is expected to be only about 0.2 of the tree amplitude.  Ways to
circumvent this difficulty include the detailed study of the isospin structure
of the $\pi \pi$ final state \cite{PP}, and the use of flavor SU(3) to
estimate penguin effects using $B \to K \pi$, where they are expected to be
dominant \cite{BPP,DGR,SilWo}.

\subsection{One determination of $\alpha$ and $\gamma$}

As an example of a way in which rate and time-dependence measurements can
shed light on both strong and weak phases, one can mention the results of
\cite{DGR} (noted above) in more detail.

(1) The decays $B^0 {\rm~or}~\bar B^0 \to \pi^+ \pi^-$ are governed by
strangeness-preserving tree and penguin amplitudes with magnitudes $|T|$ and
$|P|$, relative weak phase $\alpha$, and relative strong phase $\delta$.  By
performing time-dependent studies one can measure three quantities: the
magnitude $|A_{\pi \pi}|^2$ of the direct $B^0 \to \pi^+ \pi^-$ amplitude
$A_{\pi \pi}$, the magnitude $|\bar A_{\pi \pi}|^2$ of the direct $\bar B^0 \to
\pi^+ \pi^-$ amplitude $A_{\pi \pi}$, and an interference term ${\rm Im} (e^{2
i \beta} A_{\pi \pi} A^*_{\pi \pi})$. 

(2) The decays $B^0 \to K^+ \pi^-$ and $\bar B^0 \to K^- \pi^+$ are governed by
strangeness-changing tree and penguin amplitudes with magnitudes $|T'|$ and
$|P'|$, relative weak phase $\gamma$, and relative strong phase $\delta$
[assuming flavor SU(3)].  One relates $|T'|$ to $|T|$ using flavor SU(3) but
makes no such assumption about $|P'|$.  Two rate measurements are possible.

(3) The decays $B^\pm \to K_S \pi^\pm$ are dominated by the penguin diagram
and thus should have the same rate.  They provide information on $|P'|$. 

One thus has 6 measurements with which to determine the 6 parameters $|T|$,
$|P|$, $\alpha$, $\delta$, $|P'|$, and $\gamma$.  In general one can learn all
6, with some interesting discrete ambiguities.  Degeneracies result in some
cases, e.g., if $\Gamma(B^0 \to K^+ \pi^-) = \Gamma(\bar B^0 \to K^- \pi^+)$. 
In that case one expects also $|A_{\pi \pi}|^2 = |\bar A_{\pi \pi}|^2$ and one
has to assume something else, such as an SU(3) relation between $|P|$ and
$|P'|$, a constraint $\gamma \simeq \pi - 1.2 \alpha$ satisfied approximately
by the allowed region in Fig.~3, or both. Interesting measurements can begin to
be made with about 100 $\pi^+ \pi^-$ decays but the full power of the method
requires about 100 times more data than that. 

\section{Amplitude triangles and quadrangles}

A number of constructions of amplitude triangles and quadrangles allow one
to determine both strong and weak phases.  Measurements based on
the rates for $B^\pm \to K^\pm \pi^0,~K_S \pi^\pm,~K^\pm \eta$, and $K^\pm
\eta'$ \cite{GReta} are one example; earlier literature may be traced from
this work.  Other discussions of amplitude triangles and quadrangles and of
decays involving $\eta$ and $\eta'$ are available \cite{triquad}.

\section{REMARKS ON BARYOGENESIS}

The following discussion is an update of one given a couple of years earlier
\cite{DPFA}.  Shortly after CP violation was discovered, Sakharov \cite{Sakh}
proposed that it was one of three ingredients of any theory which sought to
explain the preponderance of baryons over antibaryons in our Universe:  (1)
violation of C and CP; (2) violation of baryon number, and (3) a period in
which the Universe was out of thermal equilibrium. 

It was pointed out by 't Hooft \cite{tH} that the electroweak theory contains
an anomaly as a result of nonperturbative effects which conserve $B - L$ but
violate $B + L$.  If a theory leads to $B - L = 0$ but $B + L \ne 0$ at some
primordial temperature $T$, the anomaly can wipe out any $B+L$ as $T$ sinks
below the electroweak scale \cite{KRS}. Thus, many models of baryogenesis
are unsuitable in practice.

Shaposhnikov and Farrar \cite{SF} have proposed that the CP violation in the
CKM sector can be communicated to a baryon asymmetry directly at the
electroweak scale.  A recent analysis of the electroweak phase transition
\cite{WB} now excludes this possibility. One proposed alternative is the
generation of nonzero $B - L$ at a high temperature, e.g., through the
generation of nonzero lepton number $L$, which is then reprocessed into nonzero
baryon number by the `t Hooft anomaly mechanism \cite{Yana}. The existence of a
baryon asymmetry, when combined with information on neutrinos, could provide a
window to a new scale of particle physics. Large Majorana masses acquired by
right-handed neutrinos would change lepton number by two units and thus would
be ideal for generating a lepton asymmetry if Sakharov's other two conditions
are met. 

One question in this scenario, besides the form of CP violation at the
lepton-number-violating scale, is how this CP violation gets communicated to
the lower mass scale at which we see CKM phases.  In two recent
models \cite{Worah,Plumacher} this occurs through higher-dimension operators
which imitate the effect of Higgs boson couplings to quarks and leptons. 

\section{BEYOND THE STANDARD MODEL}

Many models manifest their non-standard effects first in particle-antiparticle
mixing \cite{GrL,SilWol,CKLN}.  A number of tests for these effect make use of
decays of $B$ mesons.  For example, if $B^0$ and $B_s$ mixing amplitudes are
rotated by respective phases $\theta_d$ and $\theta_s$ from their
standard-model values, the rate asymmetries in $B^0 \to J/\psi K_S$ and $B_s
\to J/\psi \phi$ will probe, respectively, $- \sin(2 \beta - \theta_d)$ and
$\sin(2 \delta + \theta_s)$, where $\delta \equiv \lambda^2 \eta$.  Systematic
prescriptions for identifying the effects of the new physics using a series of
measurements are available \cite{GrL,SilWol,CKLN}. 

Neutron and electron electric dipole moments are sensitive to effects of
2-Higgs models and supersymmetry.  Limits on the neutron moment already
provide useful constraints, while limits on the electron moment are close.
For an overview of the field see \cite{Seattle}.

A lively discussion of what kaon physics can tell us about violation of CPT
(indeed, of quantum mechanics itself) is in progress \cite{CPT}. It is probably
time to improve the old test of CPT based on comparing 2 Re $\epsilon$ (as
measured in the rate asymmetry for $K_L \to \pi^\pm \ell^\mp \nu$) with what
one calculates from the measured values of $\epsilon$ and Arg $\epsilon$. 

The simplest SU(5) model \cite{GG} for unifying the strong and electroweak
interactions predicts too small an electroweak mixing angle and too short a
proton lifetime.  Proposed remedies include supersymmetry and extended gauge
structures such as SO(10) and E$_6$.  Thus, for example, if we were to
encounter new fermions in particle searches at the highest hadron and $e^+ e^-$
collider energies, we would then like to know whether they are superpartners of
gauge bosons, exotic leptons, or something else. 

Finally, I should mention a topic which, in the words of one referee, is
``not new, though not popular'':  the compositeness of quarks and leptons.
The pattern of Fig.~1 certainly looks like a level structure of a composite
system; one success of such a description would be its ability to predict
not only masses but magnitudes and phases of CKM matrix elements.  Exploratory
steps in this direction have been taken \cite{comp}.

\section{CONCLUSIONS}

I leave you with the following exercise in pattern recognition:  What familiar
pattern do you see in Fig.~6? One can re-express the pattern as shown in
Fig.~7; perhaps it suggests something at this point. Finally, when one adds
variety to the pattern, it becomes recognizable as the periodic table of the
elements (Fig.~8). 

\begin{figure}
\centerline{\epsfysize = 0.85in \epsffile {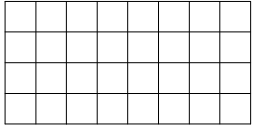}}
\caption{Part of a familiar pattern.}
\end{figure}

\begin{figure}
\centerline{\epsfysize = 0.85in \epsffile {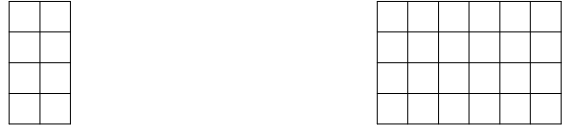}}
\caption{Part of a familiar pattern, expressed differently.}
\end{figure}

\begin{figure}
\centerline{\epsfysize = 0.85in \epsffile {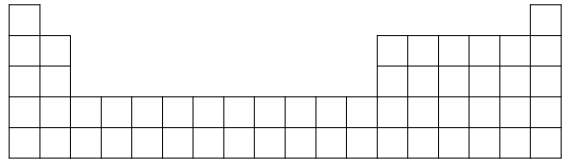}}
\caption{A larger part of a familiar pattern.}
\end{figure}

The variety of the pattern of the elementary particles laid the foundations of
the quark model and our understanding of the fundamental strong interactions.
Will there be a similar advance for quarks and leptons? The pattern of quarks
and leptons has been quite regular up to now, as if the periodic table of the
elements consisted only of rows of equal length and were missing hydrogen,
helium, the transition metals, the lanthanides, and the actinides.  Whether one
discovers superpartners of the known states, or variety such as predicted in
extended gauge structures, the new states could help us to make sense of the
pattern of the masses of the more familiar ones. 

\section{ACKNOWLEDGMENTS}

I would like to thank Jim Amundson, Isi Dunietz, Michael Gronau, Oscar
Hern\'andez,  Nahmin Horowitz, Mike Kelly, Harry Lipkin, David London, Alex
Nippe, and Sheldon Stone for enjoyable collaborations on some of the topics
mentioned here.  In addition I am grateful to Matthias Neubert, Chris
Sachrajda, Bob Sachs, Bruce Winstein, and Lincoln Wolfenstein for fruitful
discussions, and to the CERN Theory Group for its hospitality during part of
this work.  Parts of the research described here were performed at the Aspen
Center for Physics, the Fermilab Theory Group, and the Technion. This work was
supported in part by the United States Department of Energy under Grant No.~DE
FG02 90ER40560. 

\def \ajp#1#2#3{Am.~J.~Phys., #1 (#3) #2}
\def \ap#1#2#3{Ann.~Phys.~(N.Y.), #1 (#3) #2}
\def \apny#1#2#3{Ann.~Phys.~(N.Y.), #1 (#3) #2}
\def \app#1#2#3{Acta Physica Polonica, #1 (#3) #2}
\def \arnps#1#2#3{Ann.~Rev.~Nucl.~Part.~Sci., #1 (#3) #2}
\def \arns#1#2#3{Ann.~Rev.~Nucl.~Sci. #1 (#3) #2}
\def \art{and references therein}
\def \ba88{Particles and Fields 3 (Proceedings of the 1988 Banff Summer
Institute on Particles and Fields), edited by A. N. Kamal and F. C. Khanna
(World Scientific, Singapore, 1989)}
\def \baps#1#2#3{Bull.~Am.~Phys.~Soc., #1 (#3) #2}
\def \be87{Proceedings of the Workshop on High Sensitivity Beauty
Physics at Fermilab, Fermilab, Nov. 11--14, 1987, edited by A. J. Slaughter,
N. Lockyer, and M. Schmidt (Fermilab, Batavia, IL, 1988)} 
\def \cmts#1#2#3{Comments on Nucl.~and Part.~Phys., #1 (#3) #2}
\def \cn{Collaboration}
\def \corn{Lepton and Photon Interactions:  XVI International Symposium,
Ithaca, NY 1993, edited by P. Drell and D. Rubin (AIP, New York, 1994)}
\def \cp89{{\it CP Violation,} edited by C. Jarlskog (World Scientific,
Singapore, 1989)} 
\def \dpfa{The Albuquerque Meeting:  DPF 94 (Division of Particles and
Fields Meeting, American Physical Society, Albuquerque, NM, August 2--6,
1994), ed. by S. Seidel (World Scientific, River Edge, NJ, 1995)}
\def \dpff{The Fermilab Meeting -- DPF 92 (Division of Particles and
Fields Meeting, American Physical Society, Fermilab, 10--14 November, 1992),
ed. by C. H. Albright \ite~(World Scientific, Singapore, 1993)} 
\def \dpfm{The Minneapolis Meeting:  DPF 96 (Division of Particles and
Fields Meeting, American Physical Society, Minneapolis, MN, 10--15 August,
1996), to be published}
\def \dpfv{The Vancouver Meeting - Particles and Fields '91
(Division of Particles and Fields Meeting, American Physical Society,
Vancouver, Canada, Aug.~18--22, 1991), ed. by D. Axen, D. Bryman, and M. Comyn
(World Scientific, Singapore, 1992)} 
\def \efi{Enrico Fermi Institute Report No.~}
\def \fermlg{Proc.~Int.~Symp.~on Lepton and Photon Interactions at High
Energies (Fermilab, August 23--29, 1979), T. B. W. Kirk and H. D. I.
Abarbanel, eds., Fermilab, Batavia, IL (1979)}
\def \hb87{Proceeding of the 1987 International Symposium on Lepton and
Photon Interactions at High Energies, Hamburg, 1987, ed. by W. Bartel
and R. R\"uckl (Nucl.~Phys.~B, Proc. Suppl., vol. 3) (North-Holland,
Amsterdam, 1988)}
\def \ib{{\it ibid.}}
\def \ibj#1#2#3{{\it ibid.}, #1 (#3) #2}
\def \ijmpa#1#2#3{Int.~J. Mod.~Phys.~A, #1 (#3) #2}
\def \jpb#1#2#3{J. Phys., B #1 (#3) #2}
\def \jpg#1#2#3{J. Phys., G #1 (#3) #2}
\def \kdvs#1#2#3{Kong.~Danske Vid.~Selsk., Matt-fys.~Medd., #1 (#3) No.~#2}
\def \kylg{Proceedings of the International Symposium on Lepton and
Photon Interactions at High Energy, Kyoto, Aug.~19-24, 1985, edited by M.
Konuma and K. Takahashi (Kyoto Univ., Kyoto, 1985)} 
\def \latm{Lattice 1995 (Proceedings of the International Symposium on
Lattice Field Theory, Melbourne, Australia, 11--15 July 1995, T. D. Kieu,
B. H. J. McKellar, and A. Guttman, eds., North-Holland, Amsterdam, 1996}
\def \lgb{LP95: Proceedings of the International Symposium on Lepton and
Photon Interactions (IHEP), 10--15 August 1995, Beijing, People's Republic of
China, Z.-P. Zheng and H.-S. Chen, eds., World Scientific, Singapore, 1996} 
\def \lgg{International Symposium on Lepton and Photon Interactions, Geneva,
Switzerland, July, 1991}
\def \lkl87{Selected Topics in Electroweak Interactions (Proceedings of 
the Second Lake Louise Institute on New Frontiers in Particle Physics, 15--21
February, 1987), edited by J. M. Cameron \ite~(World Scientific, Singapore,
1987)}
\def \lti{lectures at this Institute}
\def \mpla #1#2#3{Mod.~Phys.~Lett., A #1 (#3) #2}
\def \nc#1#2#3{Nuovo Cim., #1 (#3) #2}
\def \np#1#2#3{Nucl.~Phys., #1 (#3) #2}
\def \npbps#1#2#3{Nucl.~Phys.~B (Proc.~Suppl.), #1 (#3) #2}
\def \oxf{Proceedings of the Oxford International Conference on
Elementary Particles 19/25 Sept.~1965, ed.~by T. R. Walsh (Chilton, Rutherford
High Energy Laboratory, 1966)}
\def \pascos{PASCOS 94 (Proceedings of the Fourth International
Symposium on Particles, Strings, and Cosmology, Syracuse University, 19--24
May 1994), ed.~by K. C. Wali (World Scientific, Singapore, 1995)}
\def \pbarp{AIP Conference Proceedings 357: 10th Topical Workshop on
Proton-Antiproton Collider Physics, Fermilab, May 1995, ed.~by R. Raja and J.
Yoh (AIP, New York, 1996)}
\def \pisma#1#2#3#4{Pis'ma Zh. Eksp. Teor. Fiz., #1 (#3) #2 [JETP Lett.,
#1 (#3) #4]} 
\def \pl#1#2#3{Phys.~Lett., #1 (#3) #2}
\def \pla#1#2#3{Phys.~Lett., A #1 (#3) #2}
\def \plb#1#2#3{Phys.~Lett., B #1 (#3) #2}
\def \ppmsj#1#2#3{Proc.~Phys.~Math.~Soc.~Jap. #1 (#3) #2}
\def \pnpp#1#2#3{Prog.~Nucl.~Part.~Phys., #1 (#3) #2}
\def \pr#1#2#3{Phys.~Rev., #1 (#3) #2}
\def \prd#1#2#3{Phys.~Rev., D #1 (#3) #2}
\def \prl#1#2#3{Phys.~Rev.~Lett., #1 (#3) #2}
\def \prp#1#2#3{Phys.~Rep., #1 (#3) #2}
\def \ptp#1#2#3{Prog.~Theor.~Phys., #1 (#3) #2}
\def \ptwaw{Plenary talk, XXVIII International Conference on High Energy
Physics, Warsaw, July 25--31, 1996}
\def \rmp#1#2#3{Rev.~Mod.~Phys., #1 (#3) #2}
\def \si90{25th International Conference on High Energy Physics, Singapore,
Aug. 2-8, 1990, Proceedings edited by K. K. Phua and Y. Yamaguchi (World
Scientific, Teaneck, N. J., 1991)}
\def \slaclg{Proceedings of the 1975 International Symposium on
Lepton and Photon Interactions at High Energies, Stanford University,
Aug.~21--27, 1975, W. T. Kirk, ed., SLAC, Stanford, CA, (1975)} 
\def \slc{Proceedings of the Salt Lake City Meeting (Division of
Particles and Fields, American Physical Society, Salt Lake City, Utah, 1987),
ed. by C. DeTar and J. S. Ball (World Scientific, Singapore, 1987)}
\def \smass{Proceedings of the 1982 DPF Summer Study on Elementary
Particle Physics and Future Facilities, Snowmass, Colorado, edited by R.
Donaldson, R. Gustafson, and F. Paige (World Scientific, Singapore, 1982)}
\def \smassa{Research Directions for the Decade (Proceedings of the
1990 DPF Snowmass Workshop), edited by E. L. Berger (World Scientific,
Singapore, 1991)}
\def \smassb{Proceedings of the Workshop on $B$ Physics at Hadron
Accelerators, Snowmass, Colorado, 21 June--2 July 1994, ed.~by P. McBride
and C. S. Mishra, Fermilab report FERMILAB-CONF-93/267 (Fermilab, Batavia, IL,
1993)} 
\def \stone{B Decays, edited by S. Stone (World Scientific, Singapore,
1994)}
\def \tasi{Testing the Standard Model (Proceedings of the 1990
Theoretical Advanced Study Institute in Elementary Particle Physics),
edited by M. Cveti\v{c} and P. Langacker (World Scientific, Singapore, 1991)}
\def \tw{this Workshop}
\def \waw{XXVIII International Conference on High Energy
Physics, Warsaw, July 25--31, 1996}
\def \yaf#1#2#3#4{Yad.~Fiz., #1 (#3) #2 [Sov.~J.~Nucl.~Phys.,~#1 (#3) #4]}
\def \zhetf#1#2#3#4#5#6{Zh.~Eksp.~Teor.~Fiz., #1 (#3) #2 [Sov.~Phys.~--
JETP, #4 (#6) #5]}
\def \zhetfl#1#2#3#4{Pis'ma Zh.~Eksp.~Teor.~Fiz., #1 (#3) #2 [JETP
Letters, #1 (#3) #4]}
\def \zp#1#2#3{Zeit.~Phys., #1 (#3) #2}
\def \zpc#1#2#3{Zeit.~Phys., C #1 (#3) #2}

\end{document}